\documentclass[10pt,a4paper]{article}
\usepackage{amsmath}
\usepackage{amsthm}
\numberwithin{equation}{section}
\usepackage{graphicx,lscape}
\begin{document}

\title{  \bf{ Studies on Regional Wealth Inequalities: \\the case of Italy}}
\author{ Marcel Ausloos$^{1,2,3}$ and  Roy Cerqueti$^{4}$}
\date{
$^{1}$School of Management, University of Leicester, \\
 University Road, Leicester, LE1 7RH,  UK \\$e$-$mail$ $address$:ma683@le.ac.uk
\\ $^{2}$eHumanities
group\footnote{Associate Researcher}$\;$, \\Royal Netherlands
Academy of Arts and Sciences, \\  Joan Muyskenweg 25, 1096 CJ
Amsterdam, The Netherlands
\\$^3$GRAPES\footnote{Group
of Researchers for Applications of Physics in Economy and Sociology}$\;$,
  rue de la Belle Jardiniere 483, \\B-4031, Angleur, Belgium \\$e$-$mail$ $address$:
marcel.ausloos@ulg.ac.be \\
$^4$University of Macerata, Department of
Economics and Law,\\  via Crescimbeni 20,   I-62100, Macerata, Italy
\\  $e$-$mail$ $address$: roy.cerqueti@unimc.it
 }

 \maketitle
 \date{submitted on Jan. 24, 2016}
\begin{abstract}
The paper contains a short review of  techniques examining regional wealth inequalities based  on recently  published research  work \cite{QQ49.15.2307RCMAstatistical,PhA421.15.187RCMAevidence,JSTAT_RCMA,SSD15RCMAseanalysisIThagio,MARCcliomQQ,EPJB87.14.261MirRCMABenfordIT}, but  also presenting unpublished features.
 The data pertains to Italy (IT),  over the period 2007-2011: the number of cities in regions, the number of inhabitants in cities and in regions, as well as  the aggregated tax income of the cities and of regions.   Frequency-size plots and cumulative distribution function plots,   scatter plots and rank-size plots are displayed.  The rank-size rule of a few cases  is   discussed.  Yearly data  of the aggregated tax income is transformed into a few indicators: the  Gini, Theil, and Herfindahl-Hirschman indices. Numerical results confirm that  IT  is  divided into  very different regional realities.  One region  is selected for a short discussion: Molise.
 A note on the "first digit Benford law" for testing data validity is presented.
\end{abstract}



\section{Introduction}
In studying geo-complexity, many variables come to mind. Beside
geo-political aspects, many considerations are social, economic and
financial ones.   One is aware that local aspects should be
distinguished from global ones. Various levels can be imagined for
discussion and modeling. Here below, the paper contains a short
review of  techniques and results examining regional wealth
inequalities based  on recently  published research  work
\cite{QQ49.15.2307RCMAstatistical,PhA421.15.187RCMAevidence,JSTAT_RCMA,SSD15RCMAseanalysisIThagio,MARCcliomQQ,EPJB87.14.261MirRCMABenfordIT},
but also presenting unpublished features. In brief, our aim is to
find correlations between long lived macro-level and shorter lived
micro-level features.

The data pertains to Italy (IT),  over the quinquennium period
2007-2011: (i) the number of cities in regions, (ii) the number of
inhabitants in cities and (iii) in regions, based on the 2011
Census, as well as  (iv)  the yearly aggregated tax income of the
cities and (v) of regions,  from 2007 till 2011, - but the ideas can
be carried forward to other countries or systems.  Rank-size plots
and cumulative distribution function plots
are displayed in Sect. \ref{sec:administration}.
Frequency size and scatter plots are to be found in the
listed references. 

In Sect. \ref{sec:ATI}, the yearly data  of the aggregated tax
income (ATI), i.e. the city contribution to the national GDP from
citizens income tax, are explored through a rank-size analysis. In
Sect . \ref{sec:financial} financial inequalities are examined: the
ATI data is transformed into a few usual indicators of wealth
distributions: the  Theil index \cite{Theil}, a Herfindahl-Hirschman
index \cite{Hirschman} and the Gini coefficient \cite{gini}.


One region has been selected for illustrating a smaller entity than
the entire country, i.e. the Molise region.  In order not to
distract from the aim,  a consideration on data validity, with
emphasis on Molise  and IT  is given at the end through Benford Law
first digit \cite{Newcomb,Benford}; see Sect. \ref{BL1}.




 \section{Administrative  distribution of cities in regions}\label{sec:administration}

As far as  2011, Italy is composed of 20 regions,  110  provinces
and 8092 municipalities, respectively. Since 2007, the number of IT
cities has been yearly evolving  to become respectively: 8101, 8094,
8094, 8092, 8092,  --  from 2007 till 2011.

In brief, several cities have thus merged into new ones, other were
phagocytized. To sum up:  13   cities became 4 ones in  two steps,
over the studied quinquennium. For completeness, let us mention
them: Campolongo Tapogliano  (UD), Ledro (TN),  Comano Terme (TN),
and Gravedona  ed  Uniti (CO).
Moreover, a few  (7) cities 
changed both province and region membership: such 7 cities moved
from PU (the province of Pesaro and Urbino) in the Marche region, to
RN (province of Rimini) in the Emilia Romagna region (Casteldelci,
Maiolo, Novafeltria, Pennabilli, San Leo, Sant' Agata Feltria,
Talamello). For completeness, notice that 228 municipalities have
changed from a  province to another one, but nevertheless remained
in the same region.


For completeness, let it be mentioned that in 2007,  the  number of
provinces was  increased from 103  (to 110) by 7 units (BT, CI, FM,
MB, OG, OT, VS)\footnote{
 see ISO code:
\textit{http://en.wikipedia.org/wiki/Provinces$\_$of$\_$Italy.}  . }
thereafter, leading to 110 provinces; 4 provinces have been
instituted by a regional law of 12 July 2001 in Sardinia and became
operative in 2005 (CI, MB, OG, OT), while BT, FM and VS have been
created on June 11th, 2004 and became operative on June 2009.

The number of regions has remained
constantly equal to 20  since 1963.
It should be here mentioned that Molise (MOL)
is the youngest Italian region, being established in 1963,
when the region "Abruzzi e Molise"   was split. Campobasso (CB) and Isernia  (IS) are the Italian provinces of Molise.

For further historical and data details, we address the reader to:
$http://www.comuni-italiani.it/regioni.html$. The above indicates
that some care had to be taken to compare entities at various times
in the quinquennium, i.e.  when necessary "data manipulation through
summation"  occurred.

\begin{figure}
  \includegraphics[width=5.8in]{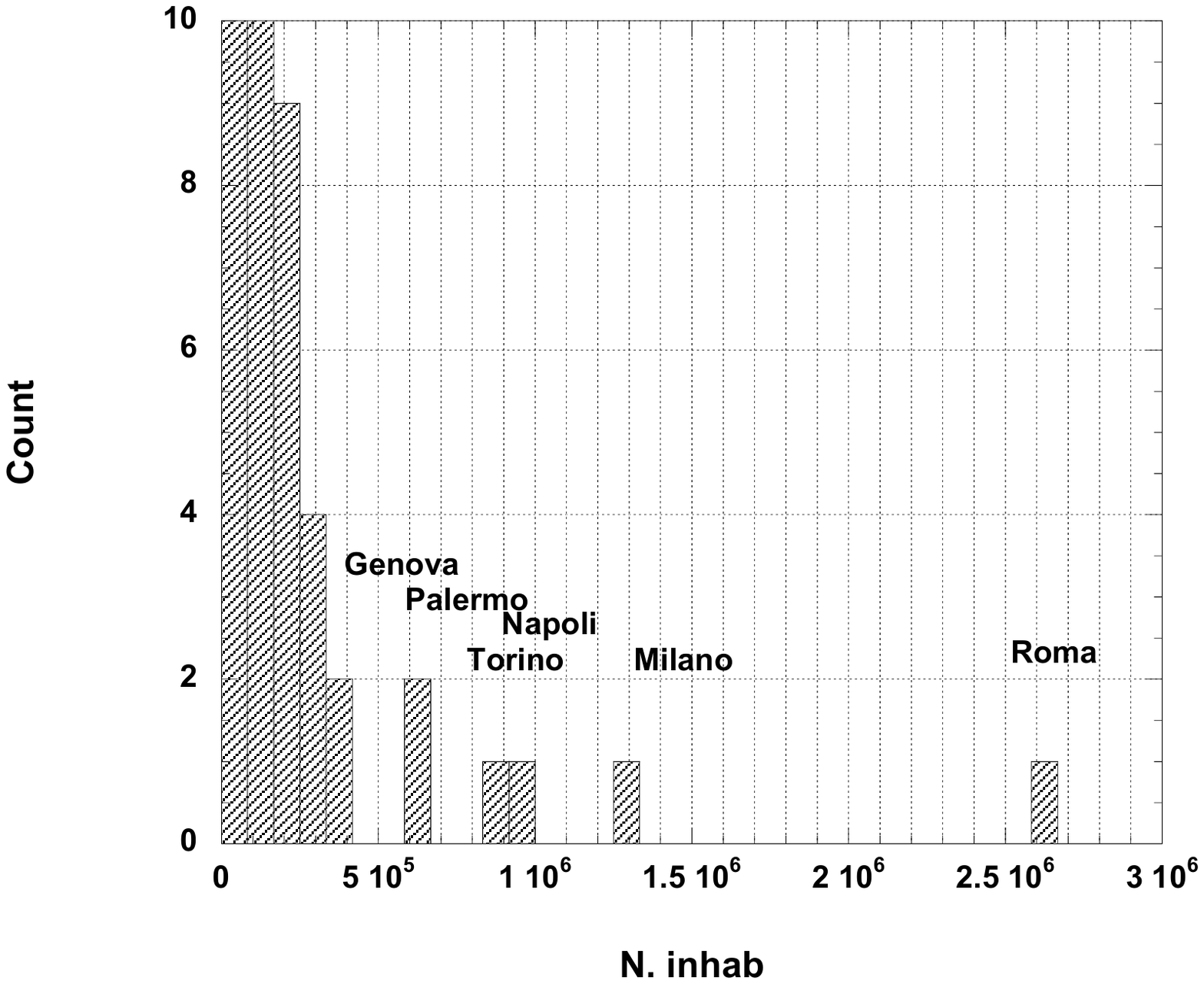}
\caption{  Histogram of the number of large cities in IT having a population of $N$ inhabitants; the $y-$ axis is truncated at a value = 10 for readability  in order to  emphasize the 6 main cities. } \label{fig:Plot 1IT Ninhablilo}
\end{figure}

\begin{figure}
\includegraphics[width=5.8in]{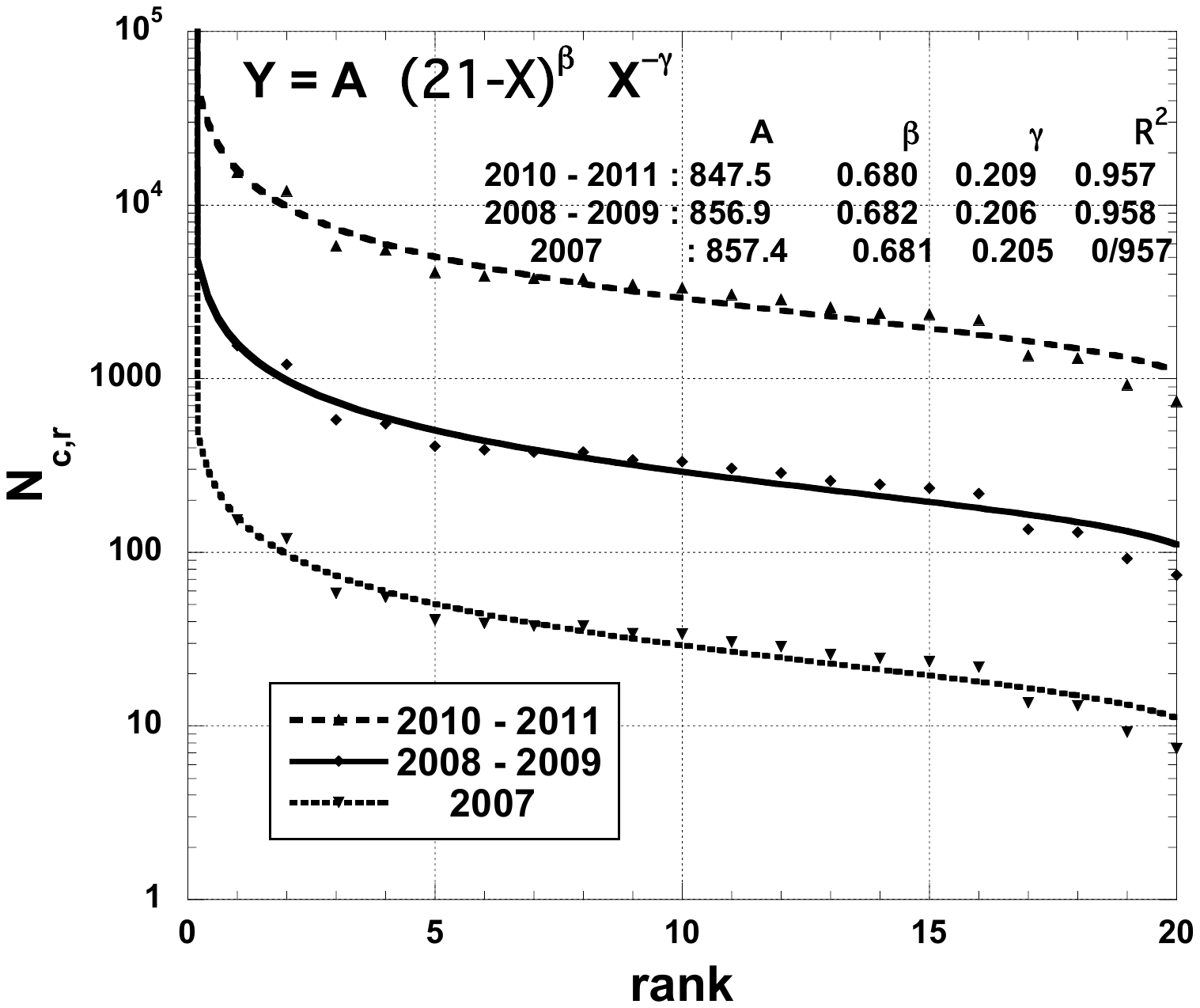}
\caption   {Display of the rank-size law of the  number ($N_{c,r}$) of cities in a given region  for the IT regions ($N$=20) in various recent years; data for 2007 and for 2010-2011 are displaced for readability;   fit values for a 3-parameter  function, as Eq. (\ref{Lavalette3a}), are  given for the undisplaced data  } \label{fig:PlotNcr0711bw3fLav3}
\end{figure}

Let us illustrate the global view of IT  cities through a histogram
presenting (counting)  the number of  cities having a population of
$N$ inhabitants, as in  Fig.\ref{fig:Plot 1IT Ninhablilo}, according
to the 2011 Census. In order to  emphasize the 6 main cities. the
$y$-axis is truncated at a value = 10 for readability. It can be
shown that these 6 cities are outliers in subsequent statistical
analysis of all pertinent data. In so doing, the notion of King (K)
and that of Vice-Roy (VR) cities can be introduced
\cite{PhA421.15.187RCMAevidence,jefferson,Sofia3a}. Such outliers
always occur, at low rank when considering rank-size plots, and much
influence classical fits with simple empirical laws (see also a
discussion below). For example, the rank-size law of the  number
($N_{c,r}$) of cities in a given region  for the IT regions ($N$=20)
in various recent years  is shown in  Fig.
\ref{fig:PlotNcr0711bw3fLav3}. The data for 2007 and for 2010-2011
are displaced for readability. The  fit  of   $N_{c,r}$, for the
(undisplayed) data,  with a 3-parameter  function
 \begin{equation} \label{Lavalette3a}
  g_3(r)= A\;   (N\;r)^{- \gamma} \;   (N-r+1)^{\beta }
\end{equation}
underlines the existence of 3 "regimes": at low, middle and high
rank, Fig. \ref{fig:PlotNcr0711bw3fLav3}.  Observe that the function
presented in Eq. (\ref{Lavalette3a}) generalizes the mere power law
$N_r = N_1 \;  r^{-\alpha} $, and takes into account the finite size
of the system and the strict vanishing of the distribution at some
maximum rank
\cite{Sofia3a,RRPh49.97.3popescu,Glottom6.03.83-93popescu,JoI1.07.155Mansilla,JQL18.11.274,MAJAQM}
.




\section{Aggregated Tax Income}\label{sec:ATI}
The analysis of rank-size for ATI confirms a situation appearing
also in other rank-size studies of other economic data. On one hand,
it was observed that the distribution of ATI between regions is
involved with time, more precisely than demographic values
\cite{QQ49.15.2307RCMAstatistical,PhA421.15.187RCMAevidence,JSTAT_RCMA,SSD15RCMAseanalysisIThagio},

On  the other hand, looking at the data,  after  calculating either
the number ($N_{i,r}$) of inhabitants in a region or the regional
ATI ($ATI_r$), i.e. the ATI sum for the relevant cities in a region,
for each year and also for the subsequent quinquennium average, the
change in regional membership  appeared to be very weakly relevant.
Thus, presenting unpublished data, consider the case of the Molise
(MOL) region.  Molise  has $N =136$ cities  (and is made of 2
provinces). A rank-size law, is hardly represented by a power law
(not shown), nor by a   simple  2-parameter free function   taking
into account the finite size of the system, as in  \begin{equation}
\label{Lavalette2} g_2(r)= \kappa_2\; \big[\frac{N\;r}{ N-r+1} \big]
^{-\chi},
\end{equation}
even removing outliers, like the King (Campobasso) and the 3
Vice-Roys (Termoli, Isernia, Venafro) cities in MOL.

\begin{figure}
\includegraphics[width=5.8in]{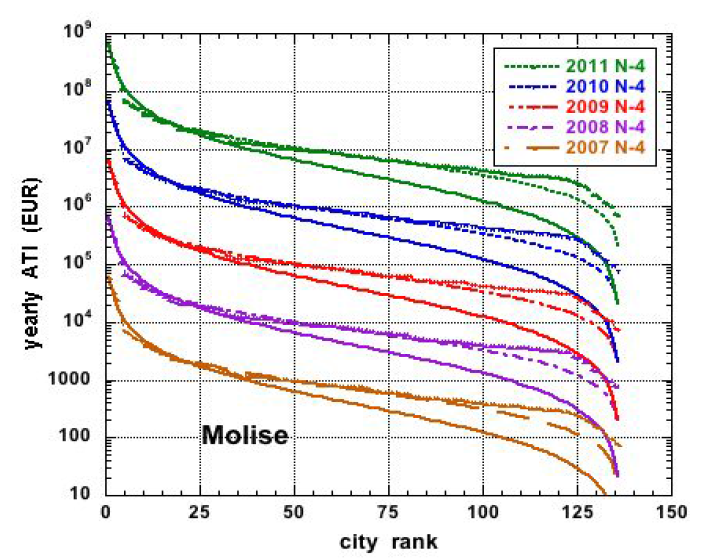}
\caption   {Displaying the rank-size law of the yearly ATI for Molise, with $N =136$   cities, but after removing the King and 3 Vice-Roy cities, in various recent years: data for 2007-08 and for 2010-2011 are displaced for readability;   fit values for a 2-parameter  function, as Eq. (\ref{Lavalette2}), are  given for the (undisplaced) data  in Table \ref {7molise}.} \label{fig:shotMoliseATI4Lav2}
\end{figure}

\begin{figure}
\includegraphics[width=5.8in]{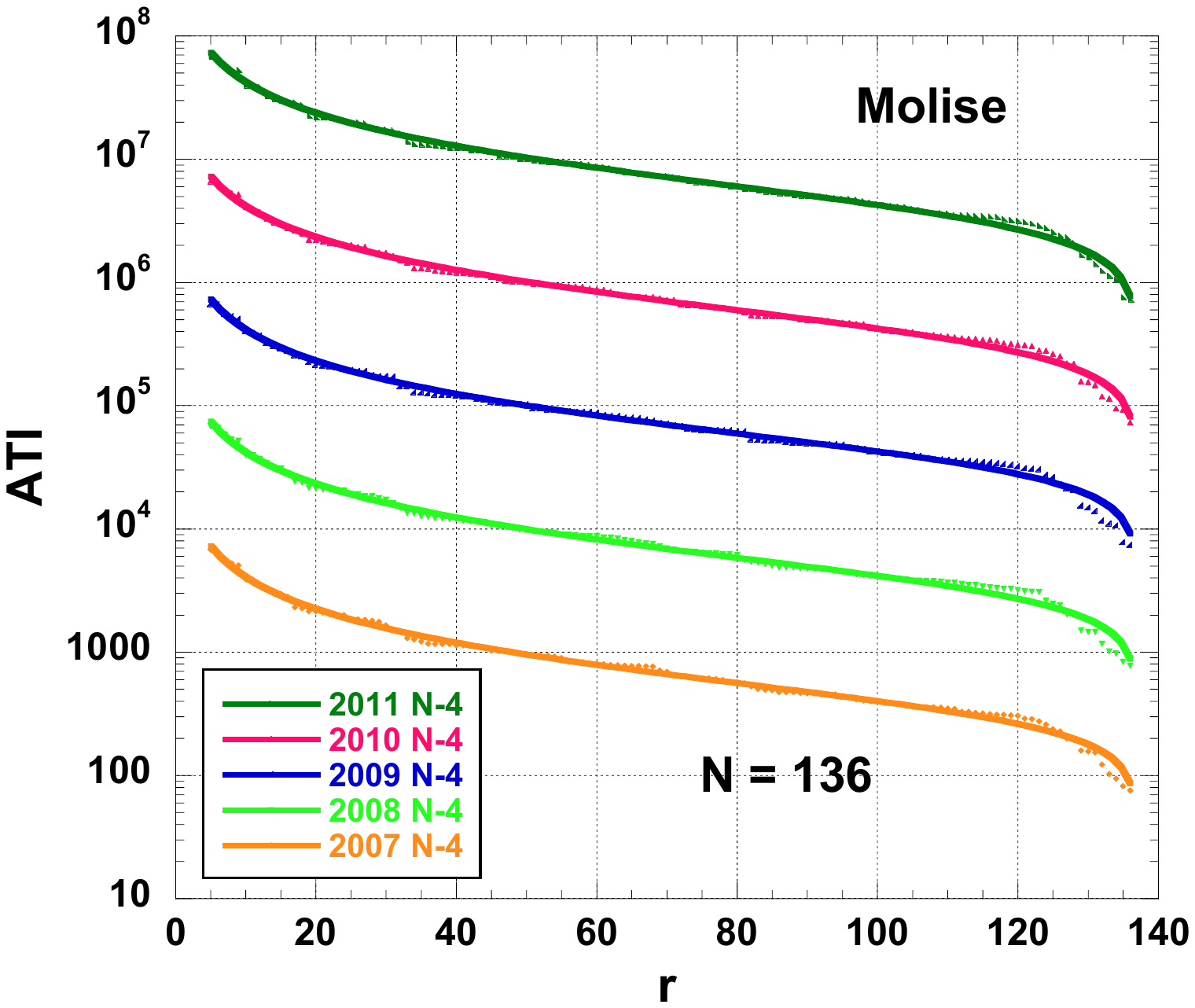}
\caption   {Displaying the rank-size law of the ATI for Molise, with $N =136$   cities, but after removing the King and 3 Vice-Roy cities, in various recent years: data for 2007-08 and for 2010-2011 are displaced for readability;   fit values for a 3-parameter  function, as Eq. (\ref{Lavalette3a}), are  given for the (undisplaced) data  in Table \ref {7molise}. } \label{fig:shotMoliseATI4Lav3}
\end{figure}

These 2-parameter free and 3-parameter free functions (Eqs.
(\ref{Lavalette2})and (\ref{Lavalette3a}, respectively), for
rank-size law research are compared in Fig.
\ref{fig:shotMoliseATI4Lav2} and Fig. \ref{fig:shotMoliseATI4Lav3}.
The various fit coefficients are given in Table \ref{7molise}.

It seems, to us and hopefully  to the readers,  that one can
conclude about the necessity of first  examining histograms of
distributions,  during the examination of ranked data,    in order
to pin point relevant outliers, on one hand, and on the other hand,
in view of  finely reproducing the empirical data through adequate
but simple functions.


  \section{Financial inequalities}\label{sec:financial}

From the  above, it appears that a rank-size law can be useful for
examining the evolution of entities in a given ensemble. Other
techniques are based on the values leading to a ranking.  These
techniques are mainly based on  a few   indicators: the  Theil index
\cite{Theil}, a  Herfindahl-Hirschman  index \cite{Hirschman} and
the Gini coefficient \cite{gini}. They stemmed from  wealth
distributions investigations and indicate (financial) inequalities
or disparities.

\subsection{Theil index}\label{theil}
The Theil index  represents one of the most common statistical tools
to measure inequality among data  \cite{Miskiewicz2008,ClippeAusloosTheil,JM40}
 Basically it represents a number which synthesizes
the degree of dispersion of an agent  in a population with respect
to a given variable (which plays the role of a measure).   The most
relevant field of application of the Theil index is represented by
the measure of income diversity. The Theil index  \cite{Theil} is
defined as:
\begin{equation}\label{Theilindexeq}
    Th =-\frac{1}{N}\sum_{i=1}^N \frac{y_i}{\sum_j y_j} \cdot
\ln\left( \frac{y_i}{\sum_j y_j} \right)
\end{equation}
where $y_i$ is the  ATI  of the $i$-th city, in our case, and the
sum $\sum_j y_j$ is the aggregation of ATI in  regions or in the
entire Italy; $N$ is the relevant number of cities when aggregating,
according to the $j$ set of interest.

Moreover, it can be easily shown that  the Theil index  can be
expressed in terms of a negative entropy
\begin{equation}\label{entropyeq}
   H = -\sum_i^N \frac{y_i}{\sum_j y_j} \cdot \ln \left(\frac{y_i}{\sum_j y_j}\right)
\end{equation}
where $\frac{y_i}{\sum_j y_j}$ is the "market share" of the $i$-th
city,  thus indicating a deviation
 from the "ideal" maximum disorder, $ln(N)$:

\begin{equation}\label{HTheq}
   H \;=\;ln (N) - Th \;\;or\;\; Th= ln(N) - H.
\end{equation}

\subsection{Herfindahl index}\label{Herfindahl}

The Herfindahl index, also known as Herfindahl-Hirschman index
(HHI), is  a measure of concentration \cite{Hirschman}.  It is
usually applied   to describe company  sizes (which measure the
concentration) with respect to the entire market:
a HHI index below 0.01 indicates a highly competitive index (from a
portfolio point of view, a low HHI index implies a very diversified
portfolio).

Adapted  to the case of ATI of cities,  HHI is an indicator of the
amount of competition  (for wealth, here) among municipalities in a
region or in the entire country. The higher the value of HHI, the
smaller the number of cities with a large value of ATI, the weaker
the competition in concurring to the formation of Italian GDP.
Formally:
\begin{equation}\label{HHIeq}
HHI =  \sum_{i\in L_{50}} \left(\frac{y_i}{\sum_j y_j} \right)^2,
\end{equation}
where $L_{50}$  is the set of the 50 largest cities in terms of ATI,
and $y_i$ is the ATI of the $i$-th city. The value 50 is
conventional, whence  HHI in Eq. (\ref{HHIeq}) is the sum of the
squares of the  market shares  of the 50 largest cities, when the
market shares are expressed as fractions.



 A normalized Herfindahl index is sometimes used and defined as:
\begin{equation}\label{NormHH}
H^{*} =  \frac{\left (HHI - 1/N \right )}{ 1-1/N }.
\end{equation}
with the appropriate $N$. 

\subsection{Gini coefficient}\label{gini}

The Gini index ($Gi$) \cite{gini}  can be viewed as a measure of the
level of fairness of a resource distribution among a group of
entities. Referring to the specific case treated here, it can be
defined through the Lorenz curve, which plots the proportion $f$ of
the total Italian ATI that is cumulatively provided by the bottom
$x$\% of the cities. If Lorenz curve is a line at 45 degrees in an $
f(x)$ plot, then there is perfect equality of ATI. The Gini
coefficient is the ratio of the area that lies between the line of
equality and the Lorenz curve over the total area under the line of
equality. A Gini coefficient of zero, of course,  expresses perfect
equality, i.e. all ATI values are the same, while a Gini coefficient
equal to one (or 100\%) expresses maximal inequality among values,
e.g.  only one city contributes to the the total Italian ATI.


A new type of display is hereby proposed. It seems of interest, for
emphasizing the structure (like the maximum position and the
corresponding percentage of relevant population) to display the data
as the difference between the Lorenz curve  ($L_{r}$) and the  line
of perfect equality in ATI:
\begin{equation}\label{ginicoeffreg}
\Delta  L_{j}= \frac{1}{N_{c,r}}[   \sum_{i=1}^{j} i\; y_{i,r}  - j]
\end{equation}
with $j \le N_{c,r}$.

\subsection{Two cases: IT and MOL}  \label{sec:2casesITMOL}

As an illustration of the  values of such indices, let us give them,
and those of the corresponding quantities of interest, in Table
\ref{TableTheilHHGiniall}, when calculated for the 5 years of the
quinquennium for IT and MOL.

It is observed that the yearly variations are rather small. The
Theil index is about 1.275. The  Gini coefficient  is $\simeq 0.75$.
The HHI is approximately given by $ 7\cdot 10^{-3}$ for IT and   $
7\cdot 10^{-2}$ for MOL. Moreover,  in each case, the indices of MOL
have a value lower than the corresponding one for IT.  The scale
effect is the greatest in HHI.  Indeed, MOL is known as one of the
poorest regions in Italy. Molise is mostly mountainous, and the
economy relies heavily on agriculture and livestock raising,
although food and garment industries are undergoing some
development. The 4 richest cities are markedly dominating in terms
of wealth, thereby leading to heterogeneity,  reflected in the
indices values.

\begin{figure}
  \includegraphics[width=5.8in]   {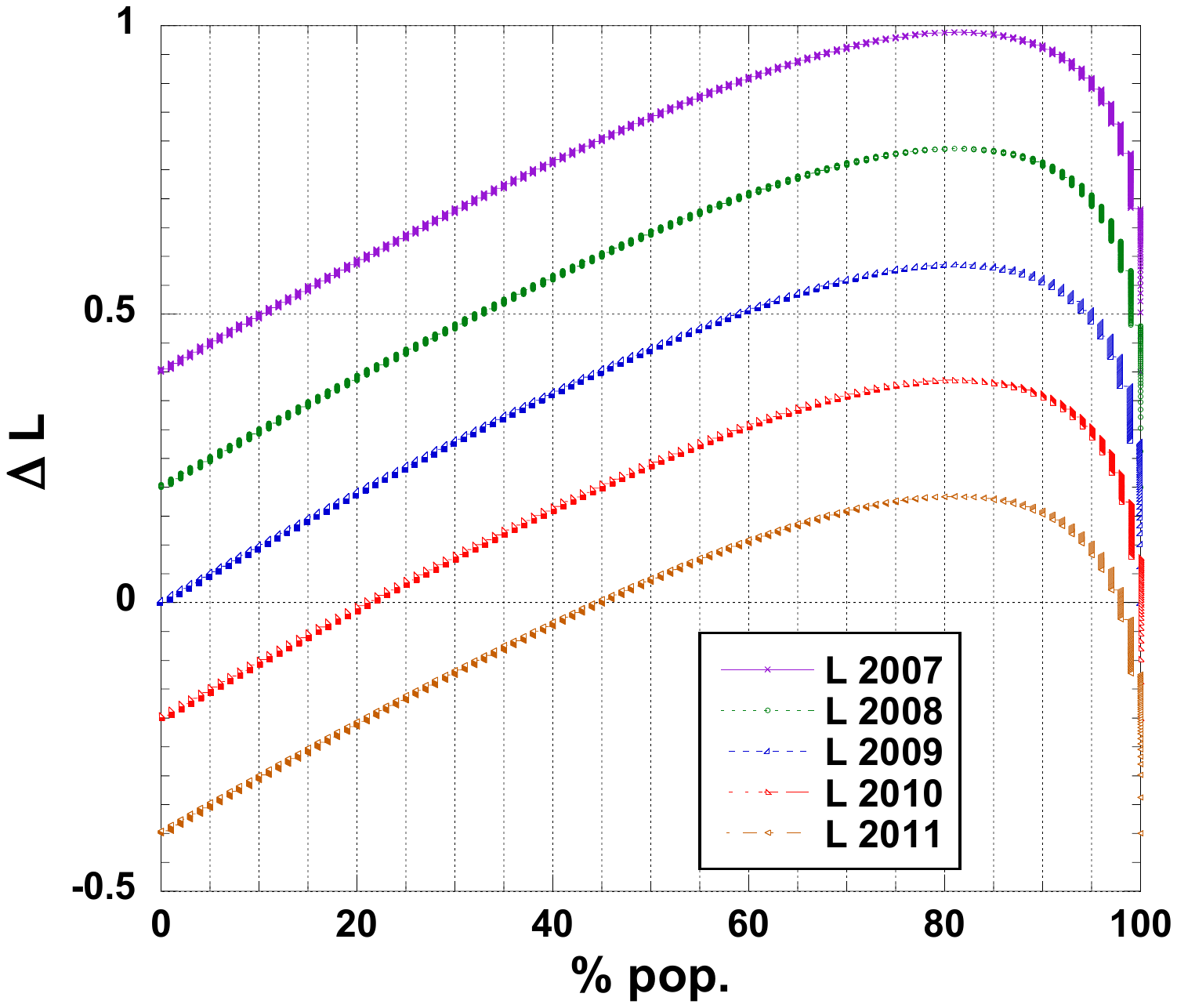}
\caption{   Another way of representing the Lorenz curve, through Eq.(\ref{ginicoeffreg}),  leading to the Gini coefficient: the figure presents the case of  the cumulated distribution of the AIT in IT 8092 cities during the 5 years of the 2007-2011 quinquennium; the data (except L009)  is displaced for readability  } \label{fig:Plot84Lpcpop0711}
\end{figure}

\begin{figure}
  \includegraphics[width=5.8in]{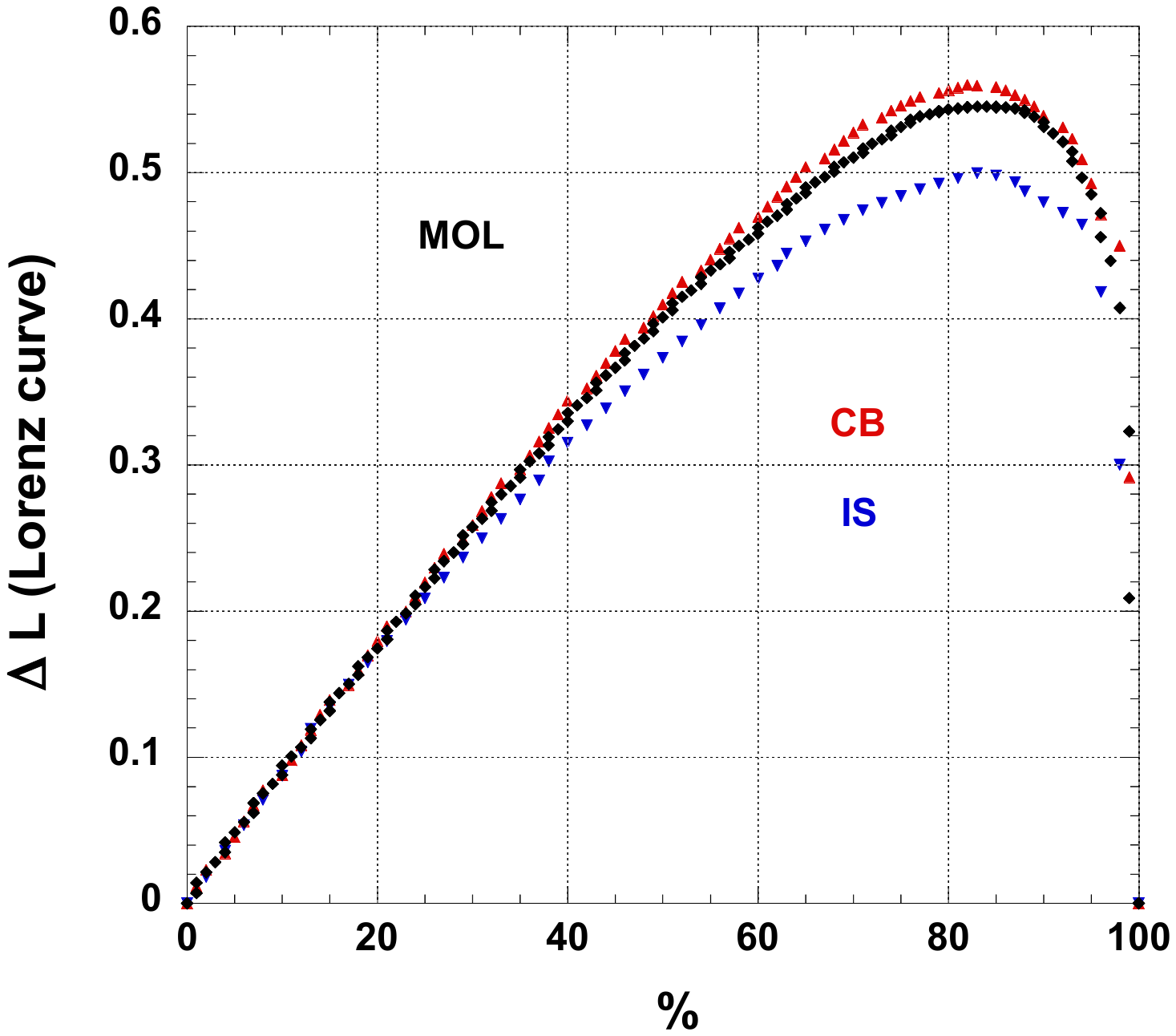}
\caption{   So called   $\Delta$ Lorenz curve,  Eq.(\ref{ginicoeffreg}),  for the cumulated distribution of the AIT in  MOL  (136 cities), but also for  IS  (52 cities) and for CB  (84 cities) cities during the 5 years of the 2007-2011 quinquennium  } \label{fig:Plot10DLorenzCBISMOL}
\end{figure}

The yearly variation of the Gini coefficient for IT has been
presented in \cite{QQ49.15.2307RCMAstatistical}.  The proposed way,
according to Eq. (\ref{ginicoeffreg}), is shown on Fig.
\ref{fig:Plot84Lpcpop0711} for IT   for the 5 years of interest.  It
can be observed that the maximum of such a curve occurs at 0.81 of
the maximum ATI. The corresponding case for Molise is shown in  Fig.
\ref{fig:Plot10DLorenzCBISMOL}, but only for the 5 year average
data, since it can be imagined that, like for IT, the Gini
coefficient of MOL hardly changed during the quinquennium; see Table
\ref{TableTheilHHGiniall} for such a justification.   However, the
maximum of the $\Delta  L_{j}$ curve  occurs at 0.83 of the maximum
ATI, - which is after finer examination slightly different from IS
and CB, and MOL. It is likely obvious, but {\it a posteriori}, that
the overall Gini coefficient and $\Delta  L_{j}$ curve are much
influenced by the larger partner in the examined set, as shown for
the case of CB  in MOL  in  Fig.  \ref{fig:Plot10DLorenzCBISMOL}.



 \section{Benford Law}\label{BL1}
Finally, one statistical tool which can serve as a
quality/reliability rapid data check    is the so-called Benford's
law  \cite{Newcomb,Benford}. First digit Benford law  for the
cumulated distribution of the  Molise 186 cities  AIT each   year of
the 2007-2011 quinquennium  is shown in Fig.
\ref{fig:BL1moliserotated}:  marked deviations are seen, in contrast
to other IT regions, as discussed in
\cite{EPJB87.14.261MirRCMABenfordIT}, even though in these some ATI
manipulation could have been expected.

 \section{Conclusion}\label{sec:conclusion}

In the main text, we have applied techniques to investigate regional
wealth  and administrative (demographically based) disparities
through recent ATI data in Italy over the  time interval 2007-2011.
The rank-size rule    is   discussed.  Yearly data  of the
aggregated tax income  when transformed into  the  Gini, Theil, and
Herfindahl-Hirschman indices has been analyzed.   One IT region has
been selected for  illustration and  discussion: Molise.
A note on the "first digit Benford law" for testing data validity
has been presented, indicating an unexpected behavior for  the
Molise region

Of course, there are other data and means  to discuss the wealth of
regions, like through   measuring   territorial dispersion rates of
small and medium-sized enterprise creation patterns
\cite{EPCNistotskaya} or as  such SME creation related to innovation
outputs \cite{STAR1}.

\section*{Acknowledgments}
\begin{figure}
  \includegraphics[width=5.8in]{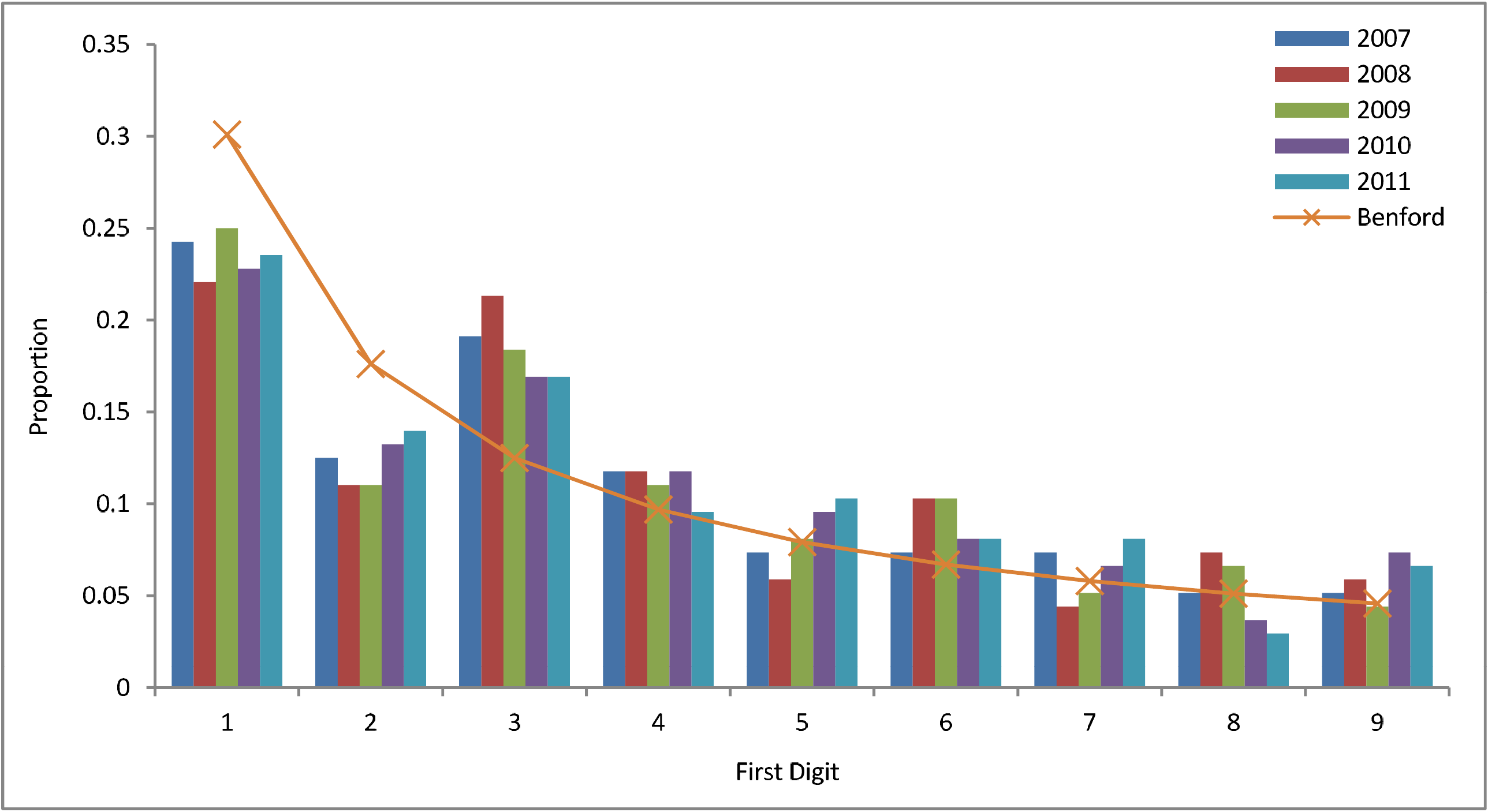}
\caption{   First digit Benford law  for the cumulated distribution of the  Molise 186 cities  AIT each   year of the 2007-2011 quinquennium  } \label{fig:BL1moliserotated}
\end{figure}

 MA thanks the FENS 2015 organizers, in particular R. Kutner and D. Grech,  for  their invitation (and as usual very warm welcome) to the Rzeszow meeting. Thanks to K. Kulakowski for providing extra logistic means. The Benford Law plot (Fig.\ref{fig:BL1moliserotated})  is courtesy of T.A. Mir.


\clearpage 
\begin{table}[htdp]
\begin{center}
\begin{tabular}{|c|c|c||c||c|c|c|c|c|}
\hline \multicolumn{3}{|c|}{ 2 param} && \multicolumn{3}{|c|}{3 param}\\ \hline
 & \multicolumn{1}{|c|}{ all}& \multicolumn{1}{|c|}{-KVR}  & &
  & \multicolumn{1}{|c|}{ all}& \multicolumn{1}{|c|}{-KVR}  \\ \hline \hline
$\kappa_2$  & 3.525$\pm$0.26&    6.443$\pm$0.113  &2007&$A$  &201.0$\pm$251.6 & 47.090$\pm$9.55  \\
\hline
$\chi$   &1.049$\pm$ 0.02 &0.730$\pm$0.007 &2007&$\gamma$ &1.079$\pm$ 0.02&0.809$\pm$0.01 \\
\hline
 &   &  &2007&$\beta$   &0.226 $\pm$ 0.25    &0.361 $\pm$0.04\\
\hline
$R^2$ &0.979  &0.990 &2007&$R^2$  &0.979 &0.993 \\
\hline\hline

$\kappa_2$  & 3.672$\pm$0.28&    6.713$\pm$0.122  &2008&$A$  &205.1$\pm$258.4 &  45.467$\pm$8.41  \\
\hline
$\chi$   &1.046$\pm$ 0.02 &0.726$\pm$0.007 &2008&$\gamma$ &1.076$\pm$ 0.02&0.795$\pm$0.01 \\
\hline
 &   &  &2008&$\beta$   &0.226 $\pm$ 0.256    &0.359 $\pm$0. 03\\
\hline
$R^2$ &0.978  &0.989 &2008&$R^2$  &0.979 &0.994 \\
\hline\hline

$\kappa_2$  & 3.539$\pm$0.26&     6.842$\pm$0.117  &2009&$A$  &318.1$\pm$351.9 &  38.2595$\pm$7.46  \\
\hline
$\chi$   &1.054$\pm$ 0.02 &0.713$\pm$0.007 &2009&$\gamma$ &1.089$\pm$ 0.02&0.783$\pm$0.01 \\
\hline
 &   &  &2009&$\beta$   &0.140 $\pm$ 0.225    &0.390 $\pm$0.04\\
\hline
$R^2$ &0.979  &0.990 &2009&$R^2$  &0.980 &0.993 \\
\hline\hline

$\kappa_2$  & 3.552$\pm$0.27&   6.929$\pm$0.115  &2010&$A$  &321.8$\pm$359.6 &  47.233$\pm$8.93  \\
\hline
$\chi$   &1.053$\pm$ 0.02 &0.709$\pm$0.007 &2010&$\gamma$ &1.087$\pm$ 0.02&0.815$\pm$0.01 \\
\hline
 &   &  &2010&$\beta$   &0.137 $\pm$ 0.227    &0.406 $\pm$0.04\\
\hline
$R^2$ &0.978  &0.990 &2010&$R^2$  &0.980 &0.993 \\
\hline\hline

$\kappa_2$  &  3.605$\pm$0.27 &     7.031$\pm$0.117  &2011&$A$  &432.8$\pm$440.0 &  35.645$\pm$7.28  \\
\hline
$\chi$   &1.053$\pm$ 0.02 &0.709$\pm$0.007 &2011&$\gamma$ &1.086 $\pm$0.02&0.780$\pm$0.01 \\
\hline
 &   &  &2011&$\beta$   &0.141 $\pm$ 0.23    &0.406 $\pm$0.04\\
\hline
$R^2$ &0.978  &0.990 &2011&$R^2$  &0.980 &0.993 \\
\hline\hline
\end{tabular}
\caption{Examples of  empirical  rank-size law fit: the case of  ATI for MOL region with $N=136$ cities,  with either 2   or 3 parameters, taking into account all data points but also when excluding the K and 3 VR   cities mentioned  in the text. Data are expressed in Euros.}\label{7molise}
\end{center}
\end{table}

\begin{table} \begin{center}
  \begin{tabular}{|c|c|c|c|c|c|c|}\hline
MOL &   2007 & 2008&2009&2010& 2011 &$<5yav>$  \\ \hline
 Entropy    &   3.6245  &   3.6314  &   3.6371  &   3.6407  &   3.6396  &3.6350\\
  Max. Entropy &   4.9127  &   4.9127  &   4.9127  &   4.9127  &   4.9127&4.9127  \\
Theil index &   1.2882  &   1.2813  &   1.2756  &   1.2719  & 1.2730&1.2777
\\ \hline
10$^2$   $HHI$   &    7.6722    &    7.6097    &    7.6336    &    7.6014    &    7.5998    & 7.6225\\
10$^2$   $H^{*}$ &    6.9883    &    6.9253    &    6.9494    &   6.9170    &   6.9153&  6.9382   \\ \hline
Gini   Coeff.    &   0.7007 &   0.6989  &   0.6959  &   0.6957  &   0.6967&0.6975 \\
\hline   \hline
whole IT   &   2007 & 2008&2009&2010& 2011&$<5yav>$  \\\hline
Entropy ($H$)&  7.2476  &   7.2603  &   7.2659  &   7.2669  &   7.2826  &   7.2650   \\
Max. Entropy &  8.9986  &   8.9986  &   8.9986  &   8.9986  &   8.9986  &   8.9986 \\
Theil index&    $\sim 1.751$   &1.7383    &1.7327    &1.7317 &1.7160
&1.7336  \\ \hline
10$^3$   $HHI$ &7.332    &   7.236   &   7.205   &   7.230   &   7.115   &   7.222  \\
10$^3$   $H^{*}$ &  7.209   &   7.113   &   7.083   &   7.107   &
6.992   &   7.099    \\ \hline
    Gini Coeff.&0.7591  &   0.7576 &    0.7566 &    0.7565 &    0.7547& 0.75685  \\
\hline
\end{tabular}
\caption{Statistical characteristics of the  ATI data
distribution as a function of time for the Molise region and for the whole Italy:  $N$= 136 and 8092, respectively. Entropy is $H$ (see
Eq. (\ref{HTheq})); Max. Entropy $\equiv ln(N)$. Theil index is taken from
Eq. (\ref{Theilindexeq}); the Herfindahl
index is $HHI$ (see Eq. (\ref{HHIeq})); the normalized Herfindahl
index is $H^{*}$ (see Eq. (\ref{NormHH})) }\label{TableTheilHHGiniall}
 \end{center}
 \end{table}

\end{document}